\documentclass[pra,aps,twocolumn,showpacs,superscriptaddress]{revtex4}
\usepackage{amsmath,graphicx,epsfig}

\begin{document}

\title{Near-zero-field nuclear magnetic resonance} 

\author{M.\ P.\ Ledbetter}\email{ledbetter@berkeley.edu}
\address{Department of Physics, University of California at
Berkeley, Berkeley, California 94720-7300}
\author{T.\ Theis}
\address{Department of Chemistry, University of California at
Berkeley, Berkeley, California 94720-3220}
\address{Materials Science Division, Lawrence Berkeley National
Laboratory, Berkeley CA 94720}
\author{J.\ W.\ Blanchard}
\address{Department of Chemistry, University of California at
Berkeley, Berkeley, California 94720-3220}
\address{Materials Science Division, Lawrence Berkeley National
Laboratory, Berkeley CA 94720}
\author{H.\ Ring}
\address{Department of Chemistry, University of California at
Berkeley, Berkeley, California 94720-3220}
\address{Materials Science Division, Lawrence Berkeley National
Laboratory, Berkeley CA 94720}
\author{P.\ Ganssle}
\address{Department of Chemistry, University of California at
Berkeley, Berkeley, California 94720-3220}
\address{Materials Science Division, Lawrence Berkeley National
Laboratory, Berkeley CA 94720}
\author{S.\ Appelt}
\address{Central Institute for Electronics, Research Center J\"ulich, D-52425 J\"ulich, Germany}
\author{B. Bl\"umich}
\address{Institute of Technical and Macromolecular Chemistry, RWTH Aachen University, 52056 Aachen, Germany}
\author{A. Pines}
\address{Department of Chemistry, University of California at
Berkeley, Berkeley, California 94720-3220}
\address{Materials Science Division, Lawrence Berkeley National
Laboratory, Berkeley CA 94720}
\author{D.\ Budker}
\address{Department of Physics, University of California at
Berkeley, Berkeley, California 94720-7300}
\address{Nuclear Science Division, Lawrence Berkeley National
Laboratory, Berkeley CA 94720}

\date{\today}


\begin{abstract}
We investigate nuclear magnetic resonance (NMR) in near-zero-field, where the Zeeman interaction can be treated as a perturbation to the electron mediated scalar interaction ($J$-coupling).  This is in stark contrast to the high field case, where heteronuclear $J$-couplings are normally treated as a small perturbation. We show that the presence of very small magnetic fields results in splitting of the zero-field NMR lines, imparting considerable additional information to the pure zero-field spectra.  Experimental results are in good agreement with first-order perturbation theory and with full numerical simulation when perturbation theory breaks down.  We present simple rules for understanding the splitting patterns in near-zero-field NMR, which can be applied to molecules with non-trivial spectra.
\end{abstract}


\maketitle

Nuclear magnetic resonance experiments are typically performed in high magnetic fields, on the order of 10 T in order to maximize chemical shifts and to achieve high nuclear spin polarization and efficient detection via inductive pickup.  The advent of various pre- or hyper-polarization schemes, and alternative methods of detection based on superconducting quantum interference devices (SQUIDs) \cite{Greenberg1998} or atomic \cite{Budker2007,Kornack2007} magnetometers has enabled NMR experiments in very low ($\approx$earth's field) and even zero magnetic field, generating significant experimental \cite{McDermott2002,Trabesinger2004,Appelt2005,Savukov2005,Xu2006,Appelt2006,
Robinson2006,Savukov2007,Appelt2007,Ledbetter2008,Ledbetter2009,Appelt2010,Theis2011} and theoretical interest \cite{Appelt2007,Appelt2010PRA,Kervern2011}. Low-field NMR carries the advantage of providing high absolute field homogeneity, yielding narrow resonance lines and accurate determination of coupling parameters \cite{Appelt2006,Ledbetter2009}.  Further, elimination of cryogenically cooled superconducting magnets facilitates the development of portable devices for chemical analysis and imaging. In this regard, atomic magnetometers are an ideal tool because, in contrast to SQUIDs, they do not require cryogenic cooling. Recent work using atomic magnetometers to detect NMR was performed at zero field, in part, because of the need to match the resonance frequencies of the nuclear spins and the magnetometer's alkali spins, which have very different gyromagnetic ratios \cite{Ledbetter2009,Theis2011}. It has been pointed out that zero-field NMR leaves some ambiguity in determination of chemical groups, and that this ambiguity can be removed by application of small magnetic fields \cite{Appelt2010PRA}.

Here, we examine, experimentally and theoretically, the effects of small magnetic fields in near-zero-field (NZF) NMR.  We show that application of weak magnetic fields results in splitting of the zero-field (ZF) lines, restoring information about gyromagnetic ratios that is lost in ZF NMR. In the regime where the Zeeman effect can be treated as a perturbation, we observe high-resolution spectra with easy-to-understand splitting patterns that are in good qualitative and quantitative agreement with first-order perturbation theory. This work represents the first observation of NMR under such conditions, forming the basis for a new type of NMR spectroscopy that serves as a complement to high-field NMR, where heteronuclear couplings are almost always treated as a small perturbation to the much larger Zeeman interaction.  We also examine the case in which the Zeeman energies are comparable to the $J$-coupling energies, resulting in spectra of maximal complexity.

The Hamiltonian in the presence of $J$-couplings and a magnetic field is
\begin{equation}\label{Eq: Hamiltonian}
    H = \hbar \sum_{j;k>j}J_{jk} \mathbf{I}_j\cdot\mathbf{I}_k - \hbar \sum_j \gamma_j \mathbf{I}_j\cdot\mathbf{B}.
\end{equation}
Here $\mathbf{I}_j$ represent both like and unlike spins with gyromagnetic ratio $\gamma_j$ and $J_{jk}$ is the scalar coupling between spins $j$ and $k$.  In the absence of magnetic fields, the spherical symmetry of the Hamiltonian dictates that eigenstates $|\phi_a\rangle$ are also eigenstates of $\mathbf{f}^2$ and $f_z$, where $\mathbf{f}$ is the total angular momentum $\mathbf{f}=\sum_j\mathbf{I}_j$, with energy $E_a$, and degeneracy $2f+1$.  Application of a magnetic field $B_z$ lifts this degeneracy, splitting the ZF NMR lines.

We first examine the effects of very small magnetic fields on a ${\rm ^{13}CH}_N$ system, with $N$ equivalent protons, using perturbation theory.  In zero field, the unperturbed energy levels are given by $E(f,k)=J/2[f(f+1)-k(k+1)-s(s+1)]$, \cite{Ledbetter2009} where $k = 1/2,1,3/2...$ are the possible spin quantum numbers of the operator $\mathbf{k}$ describing the sum of the equivalent proton spins, and $s=1/2$ is the spin quantum number associated with the operator  $\mathbf{s}$, representing the ${\rm ^{13}C}$ spin. To first order in $B_z$, eigenstates are those of the unperturbed Hamiltonian, and Zeeman shifts of the eigenvalues can be read from the diagonal matrix elements of the Zeeman perturbation.  One finds:
\begin{eqnarray}
\nonumber \Delta E(f,k,m_f)=-\langle f m_f | B_z (\gamma_h k_z+\gamma_c s_z) | f m_f \rangle\\
  = -B_z\sum_{m_k,m_s}\langle k s m_k m_s|f m_f\rangle^2 (\gamma_h m_k+\gamma_s m_s).\label{Eq:deltaE}
\end{eqnarray}
Here $\gamma_h$ and $\gamma_c$ are the proton and ${\rm ^{13}C}$ gyromagnetic ratios, and $\langle k s m_k m_s| f m\rangle$ are the Clebsch-Gordan coefficients. The observable in our experiment is the total $x$ magnetization, $M_x(t) \propto {\rm Tr} \rho(t)\sum_{j}I_{jx}\gamma_j$, where $\rho(t)$ is the time dependent density matrix.  Writing $I_{jx}$ in terms of the raising and lowering operators, we obtain selection rules for observable coherences: $\Delta f = 0,\pm 1$ and $\Delta m_f = \pm 1$, valid in the limit where $|\gamma_jB|<<|J|$.  In the case at hand with $N$ equivalent protons, there is an additional selection rule, $\Delta k = 0$, since, in the absence of chemical shifts, the Hamiltonian commutes with $\mathbf{k}^2$.

\begin{figure}
  \includegraphics[width=3.4 in]{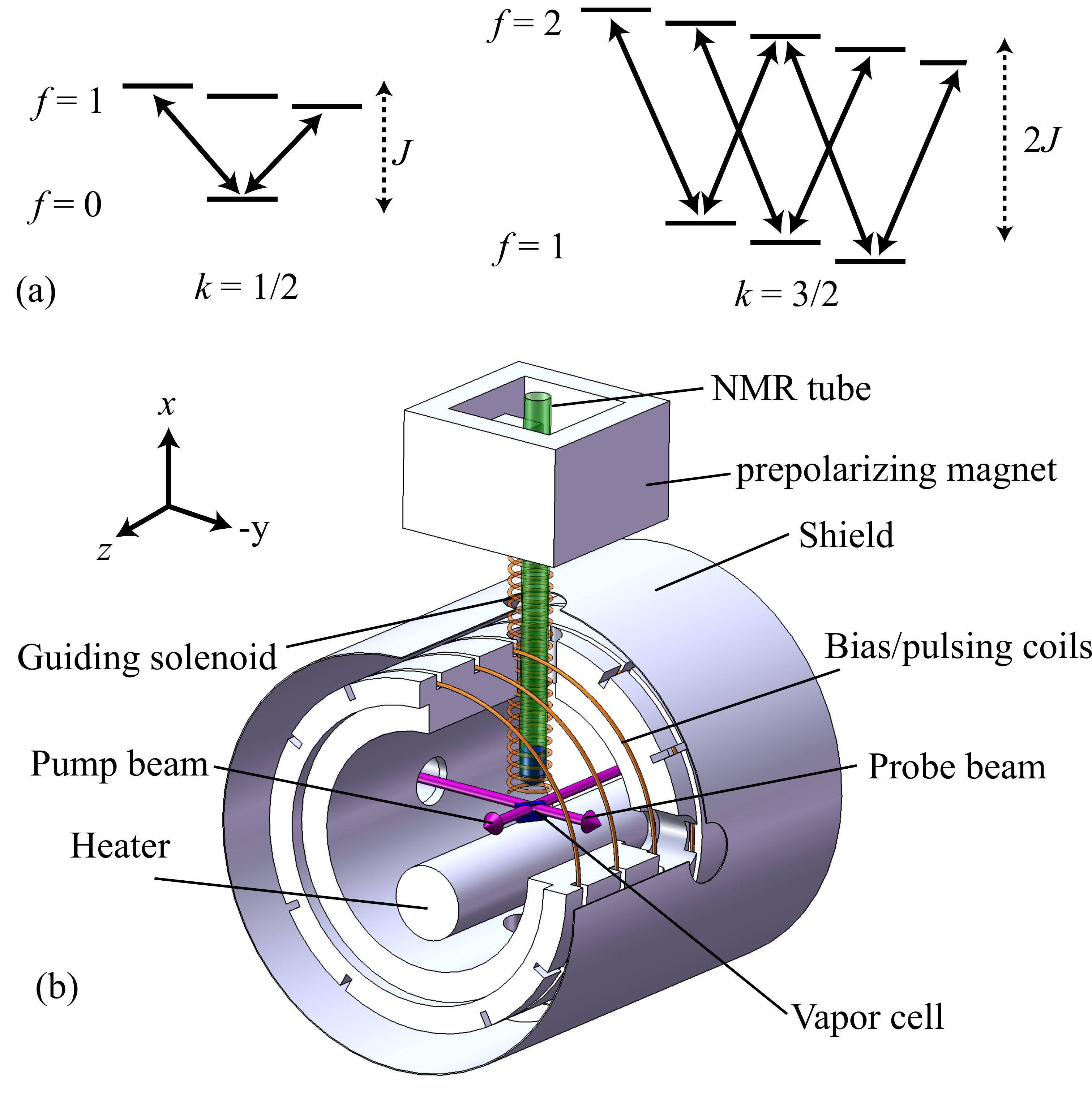}\\
  \caption{(color online) (a) Energy levels for a ${\rm ^{13}CH_3}$ group.  Energy levels for a ${\rm ^{13}CH}$ group are given by the manifold on the left.  (b) Experimental setup for near-zero-field spectroscopy, described in the text.
}\label{Fig:Apparatus}
\end{figure}

Experimentally, we examine the case of $N = 1$ and $N=3$.  In the former case, $k=1/2$, the zero-field levels are a singlet with $f=0$ and a triplet with $f = 1$. In the presence of a small magnetic field, the singlet level is unperturbed, while the triplet levels split, as shown by the manifolds on the left of Fig. \ref{Fig:Apparatus}(a).  In the following, $\nu_{f,m_f}^{f',m_f'}$ denotes the frequency of transitions between the states $|f,m_f\rangle$ and $|f',m_f'\rangle$. Employing Eq. \eqref{Eq:deltaE} and the selection rules, one finds a single line for transitions with $\Delta f = 0$ between states with $f=1$, and a doublet for transitions with $\Delta f = \pm 1$ between states with $f = 1$ and $f = 0$:
\begin{eqnarray}
\nu^{1,m_f\pm 1}_{1,m_f} &=& B_z(\gamma_h+\gamma_c)/2,\label{Eq:zero11k12}\\
\nu^{1,\pm 1}_{0,0}&=&J\pm B_z(\gamma_h+\gamma_c)/2.\label{Eq:doubletfreqs}
\end{eqnarray}

For the case of $N=3$, $k$ is either 1/2 or 3/2. The $k=1/2$ transition frequencies are given by Eq. \eqref{Eq:doubletfreqs}. The $k=3/2$ manifolds are shown on the right of Fig. \ref{Fig:Apparatus}(a), and coherences between $|f=1,m_f\rangle$ and $|f=2,m_f\pm 1\rangle$ occur at frequencies given by
\begin{equation}\label{Eq:sextetfreqs}
    \nu^{2,m_f\pm 1}_{1,m_f}=2J+m_f\frac{B_z}{4}(-7\gamma_h+6\gamma_c)\pm \frac{B_z}{4}(3\gamma_h+\gamma_c).
\end{equation}
There are two additional transitions for states with $k=3/2$ with $\Delta f = 0$ that occur near zero frequency,
\begin{eqnarray}
  \nu^{2,m_f\pm 1}_{2,m_f} &=& (3\gamma_h+\gamma_c)B_z/4; k=3/2\label{Eq:zero22k32},\\
  \nu^{1,m_f\pm 1}_{1,m_f} &=& (5\gamma_h-\gamma_c)B_z/4; k = 3/2\label{Eq:zero11k32}.
\end{eqnarray}
Equations. \eqref{Eq:zero11k12}-\eqref{Eq:zero11k32} constitute a set of
eleven transitions, three appearing near zero frequency, two near $J$, and six near $2J$, representing the NZF NMR spectrum of a ${\rm ^{13}CH_3}$ group. These calculations are discussed in more depth in the Supplementary Information, and in Ref. \cite{Appelt2010PRA}.

We now make two observations: (1) Even in more complex molecules with additional non-equivalent spins, the zero-field eigenstates are also those of $\mathbf{f}^2$ and $f_z$. Therefore, the NZF splitting patterns can be used to identify the angular momenta of the states involved in the zero-field transitions: Transitions between levels with $f=0$ and 1 will produce doublets, transitions between levels with $f=1$ and 2 will produce a multiplet with six lines, and so on.  (2) The selection rules presented here break down as the magnetic field becomes large enough to produce significant mixing of the zero-field eigenstates.  Reference \cite{Appelt2010PRA} shows theoretically that the maximum number of lines for a ${\rm ^{13}CH}_N$ group is $(N+1)^2$, most clearly visible when $|(\gamma_h+\gamma_c)B_z|\approx J$.

Experiments were performed using an apparatus similar to that of Refs. \cite{Ledbetter2009,Theis2011} and depicted in Fig. \ref{Fig:Apparatus}.  Samples (typically $\approx {\rm 200~\mu L}$) were contained in a 5 mm NMR tube, and pneumatically shuttled between a 1.8 T prepolarizing magnet and a magnetically shielded enclosure, housing a microfabricated $^{87}{\rm Rb}$ vapor cell, the central component of the atomic magnetometer. 
The cell is optically pumped by $z$-directed, circularly polarized laser light, tuned to the center of the D1 transition, and probed by $y$-directed, linearly polarized light, tuned about 100 GHz to the blue of the D1 transition.  Optical rotation of the probe light is monitored by a balanced polarimeter. Bias fields and DC pulses of magnetic field, used to excite NMR spin coherences, are applied via a set of coils.  At zero field, the magnetometer is primarily sensitive to fields in the $x$ direction with noise floor of about $40-50~{\rm fT/\sqrt{Hz}}$.  As the bias field is increased, the magnetometer response moves to higher frequencies, compromising the low-frequency sensitivity by about a factor of 5 for $B_z=3~{\rm mG}$.
To maintain a quantization axis during transit of the sample, a solenoid provides a ``guiding" field.  The guiding field is turned off suddenly prior to acquisition of data, and a pulse applied in the $z$-direction with area such that the proton spins rotate through $\approx 4\pi$ and the carbon spins rotate through $\approx \pi$ (about 480~${\rm \mu s}$), maximizing the amplitude of zero-field signals. 

\begin{figure}
  \includegraphics[width=3.3 in]{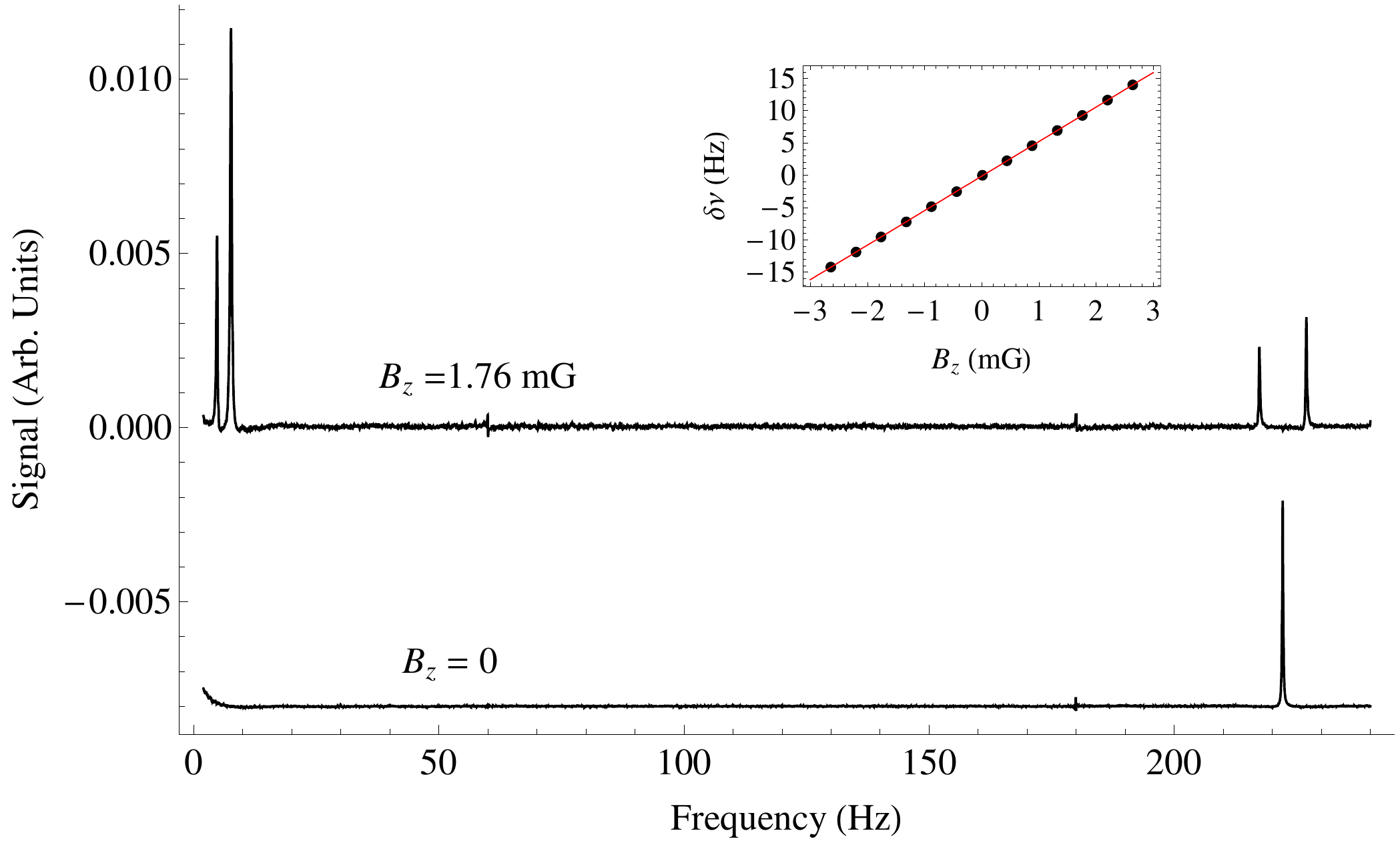}\\
  \caption{Spectra for ${\rm ^{13}C}$ labeled formic acid, ${\rm H^{13}COOH}$, in the indicated magnetic fields.  The spectra are the result of averaging eight transients.  The inset shows the splitting of the two lines centered about $J$ as a function of magnetic field.} \label{Fig:Formicacid}
\end{figure}

ZF and NZF spectra for formic acid (${\rm H^{13}COOH}$) are shown in Fig. \ref{Fig:Formicacid}. The ZF spectrum consists of a single line at $J=222~{\rm Hz}$, as well as a DC component, suppressed here for clarity.  The NZF spectrum arising from the ${\rm ^{13}CH}$ group is as discussed above: a doublet with frequencies $J\pm B_z(\gamma_h+\gamma_c)/2$ and an additional line at $B_z(\gamma_h+\gamma_c)/2\approx 4.7~{\rm Hz}$. The large peak at 7.5 Hz corresponds to the uncoupled OH group.  The asymmetry in the doublet centered about $J$, reproduced by a full numerical calculation, is due to higher-order corrections to the eigenstates.  The peaks are well described by Lorentzians, with half-width at half-maximum $\approx 0.1~{\rm Hz}$, and the locations of the peaks can be determined with an uncertainty of about 1 mHz. The inset shows the splitting of the line at $J$ as a function of magnetic field, 
displaying a linear dependence.  The slope is in agreement with that predicted by Eq. \eqref{Eq:doubletfreqs}, $(\gamma_h+\gamma_c)$, at the level of about 0.1\%.

\begin{figure}
  \includegraphics[width=3.3 in]{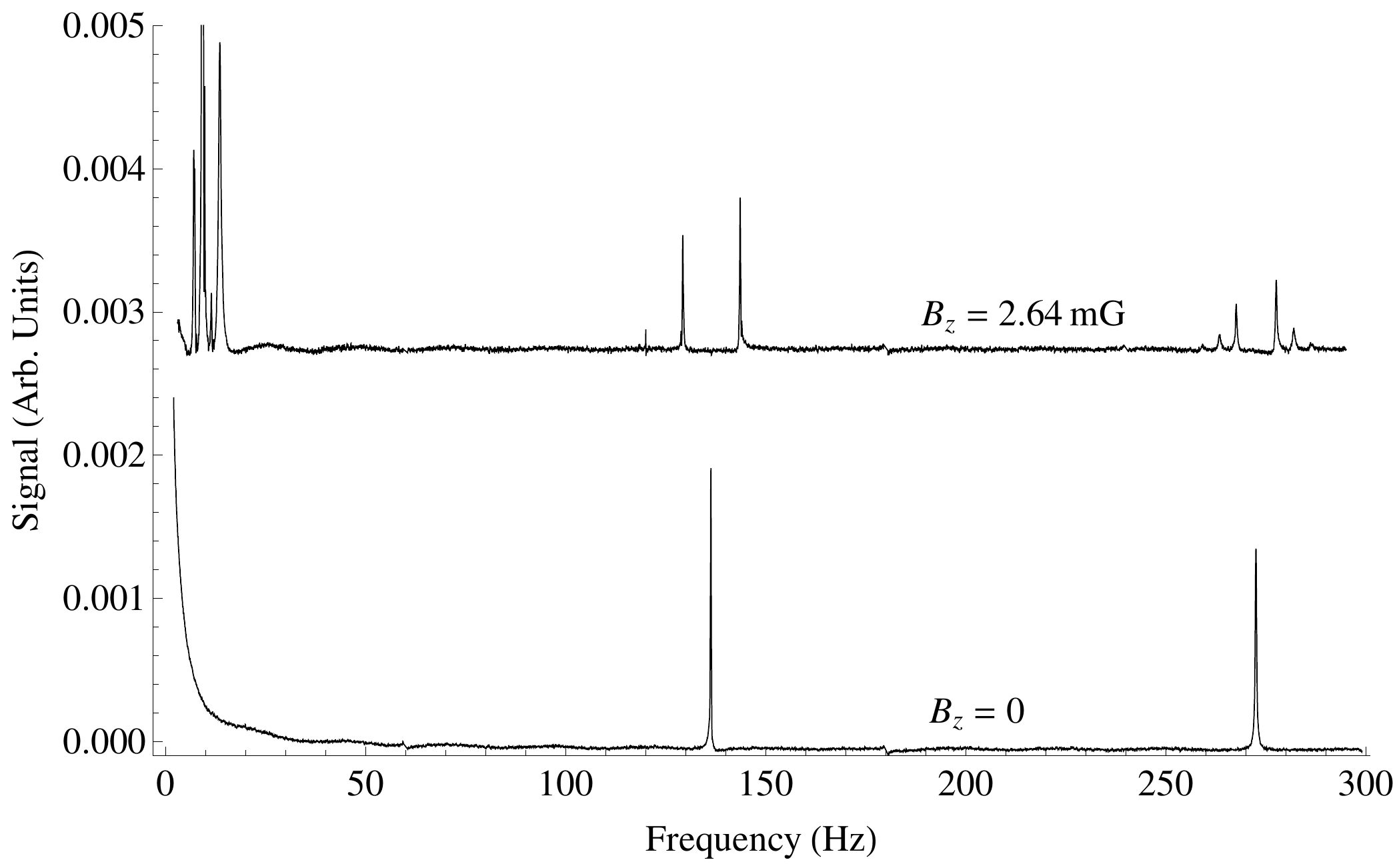}\\
  \caption{Spectra for singly labeled acetonitrile-2, ${\rm ^{13}CH_3CN}$ in zero-field and in a field of 2.64 mG.  The positions of all peaks are well described by Eqs. \eqref{Eq:zero11k12}-\eqref{Eq:zero11k32}.} \label{Fig:acetonitrile}
\end{figure}

To illustrate the case of a ${\rm ^{13}CH_3}$ system, ZF and NZF spectra for acetonitrile-2 (${\rm ^{13}CH_3CN}$) are shown in Fig \ref{Fig:acetonitrile}.  For $B_z=0$, the spectrum consists of a zero-frequency peak, a peak at $J$, and a peak at $2J$.  Application of a magnetic field splits the zero-frequency peak into three lines, whose frequencies are given by Eqs. \eqref{Eq:zero11k12},\eqref{Eq:zero22k32}, and \eqref{Eq:zero11k32}. The smallest peak at 11.2 Hz corresponds to an uncoupled proton due to an unknown solvent in the sample.  The line at $J$ splits into a doublet, whose frequencies are given by Eq. \eqref{Eq:doubletfreqs}, and the line at $2J$ splits into six lines, whose frequencies are given by Eq. \eqref{Eq:sextetfreqs}.  The splitting of the lines at $J$ and $2J$ clearly reveals the degeneracy of the zero-field levels.  As with the formic acid spectrum, there is some asymmetry present in the multiplets centered about $J$ and $2J$, which is reproduced by numerical simulation.  Nevertheless, the relative amplitudes of the lines centered about $2J$ are roughly in the ratio 1:3:6:6:3:1 as expected from first-order perturbation theory (see Supplementary Information). 

\begin{figure}
  \includegraphics[width=3.4 in]{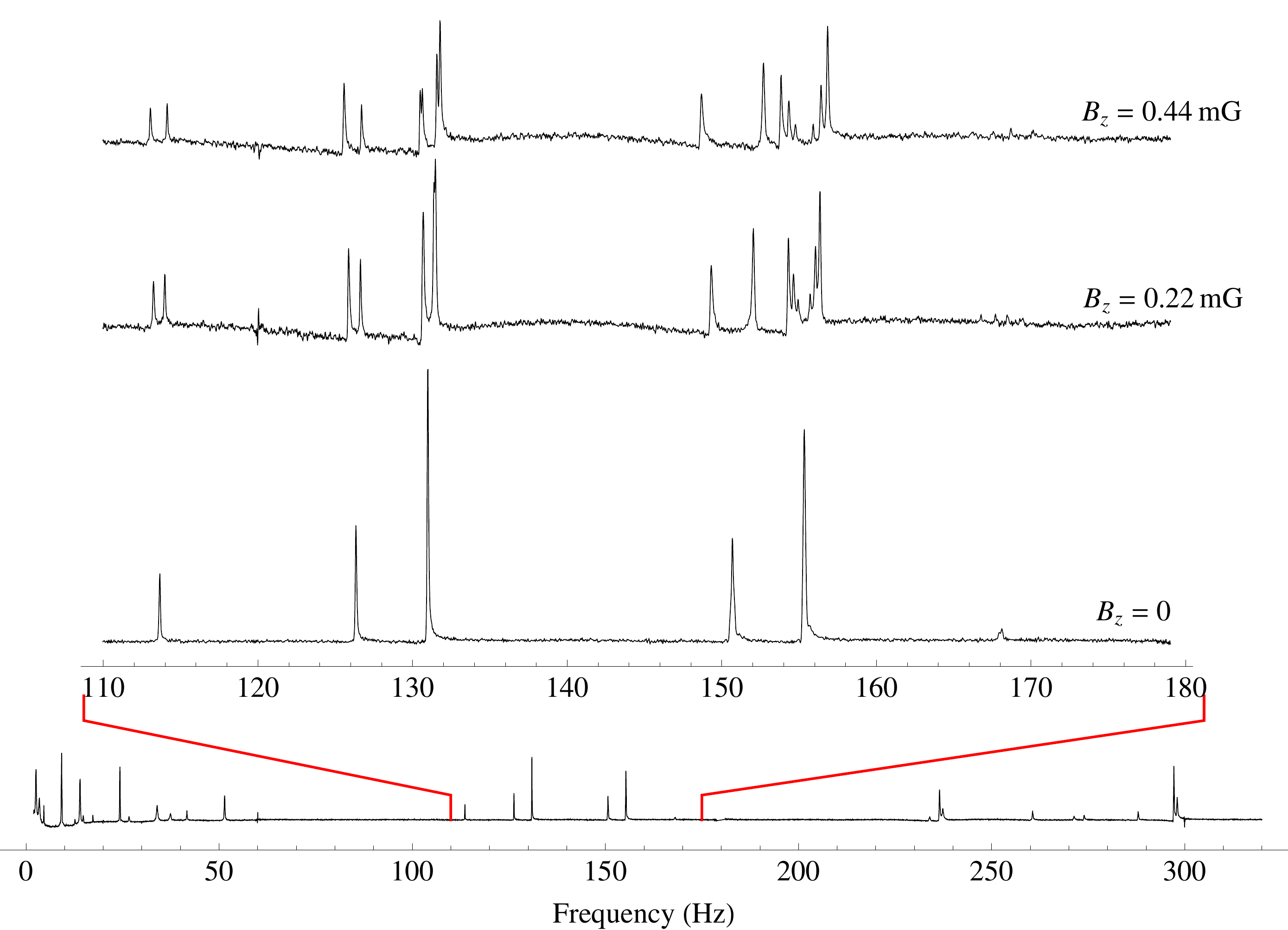}\\
  \caption{Effects of small magnetic fields on fully labeled acetonitrile ${\rm ^{13}CH_3^{13}C^{15}N}$.  The bottom trace shows the entire zero-field spectrum.  The upper traces show an expanded view of the central part of the zero-field spectrum, as well as the spectra in the indicated finite fields.}\label{Fig:fullylabeledacetontrile}
\end{figure}

To illustrate the utility of NZF NMR, we examine the case of fully labeled acetonitrile (${\rm ^{13}CH_3\thinspace^{13}C^{15}N}$).  The zero field spectrum is shown in the bottom trace of Fig. \ref{Fig:fullylabeledacetontrile}.  It is not immediately clear which lines correspond to which zero-field transitions.  An expanded view of the zero-field spectrum in the range of 110 to 180 Hz is provided and compared to the spectrum obtained in the indicated finite magnetic fields.  We see the appearance of doublets centered at 114, 126, and 151 Hz, indicating that these transitions occur between manifolds with $f=0$ and $f=1$.  It is interesting to note that these doublets display different splittings due to differences in the Land\'e $g$ factors for the different manifolds involved in these transitions. The line at 131 Hz splits first into a doublet, which split into a pair of doublets.  One can show that such a splitting pattern arises for a $f=1\leftrightarrow f=1$ (see Supplementary Information). The small zero-field peak at 168 Hz splits into four lines, barely above the noise, indicating an additional $f=1\leftrightarrow f=1$ transitions. Finally, the zero-field peak at 155.5 Hz splits into a sextet indicating the transition is $f=1\leftrightarrow f=2$.  The six lines in this multiplet appear ``inside-out" compared to the six line multiplet observed at $2J$ in 2-acetonitrile due to a reversal in relative magnitude of the Land\'e $g$ factor.

The multiplicity of the peaks in this part of the spectrum can be understood as follows: Suppose we start with a ${\rm ^{13}CH_3}$ group, and confine our attention to the $1\leftrightarrow 0$ transition with total proton spin = 1/2, yielding transitions in the neighborhood of $^1J_{\rm CH}$.  Addition of the second ${\rm ^{13}C}$ splits these levels:  $f=1$ splits to 3/2, 1/2 manifolds, and $f=0$ manifolds splits to 1/2.  Addition of the ${\rm ^{15}N}$ splits these so we now have $f_a=2$ or 1, $f_b = 1$ or 0, and $f_c = 1$ or 0. For now, we ignore transitions between $f_a \leftrightarrow f_b$ because they occur at low frequency. Employing the $\Delta f=1$ rule we expect three $1\leftrightarrow 0$ transitions, producing doublets: $f_a=1 \leftrightarrow f_c=0$, $f_b=1 \leftrightarrow f_c=0$, and $f_b=0 \leftrightarrow f_c=1$.   Transitions between $f_a=2\leftrightarrow f_c = 1$ yields a multiplet with six lines, and transitions with $\Delta f = 0$ between $f_a = 1 \leftrightarrow f_c = 1$ and between $f_b = 1 \leftrightarrow f_c = 1$ yield multiplets with four lines.  More details are presented in the Supplementary Information.

\begin{figure}
  \includegraphics[width=3.5 in]{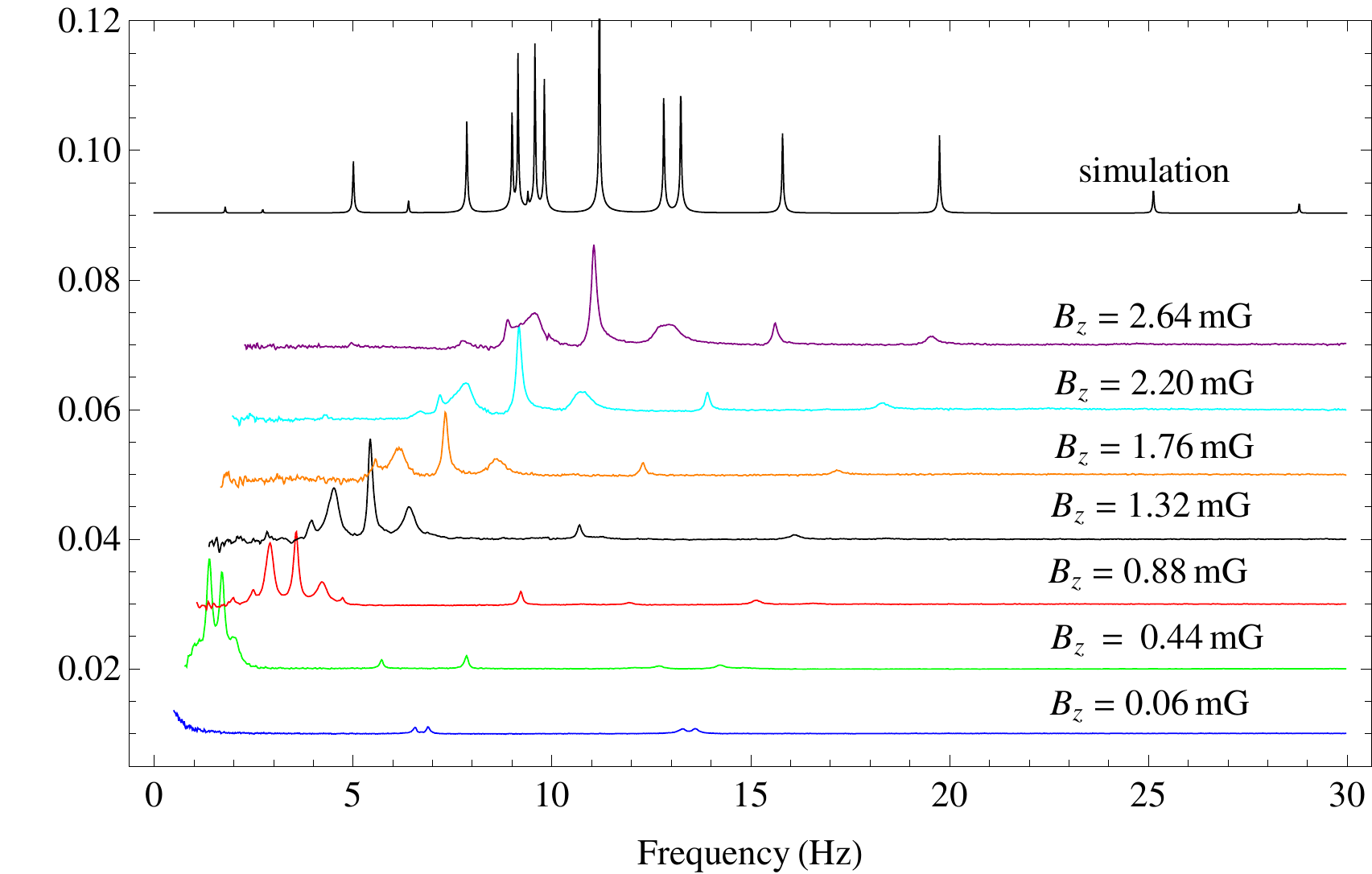}\\
  \caption{Experimental spectra for 1-acetic acid, (${\rm CH_3\thinspace^{13}COOH}$) in the indicated magnetic fields. The smooth curve at the top of the plot presents the result of a full numerical simulation with high resolution.} \label{Fig:aceticacid}
\end{figure}

In systems with small couplings,  such as 1-acetic acid (${\rm CH_3\thinspace^{13}COOH}$) which has a two-bond coupling, $^2J_{\rm CH}=6.8~{\rm Hz}$, it is possible to explore the regime in which the Zeeman interaction is comparable to the $J$-coupling.  Figure \ref{Fig:aceticacid} shows experimental spectra for 1-acetic acid for the indicated magnetic fields.  The large peak that does not split is due to the uncoupled OH group, while the rest of the spectrum corresponds to the ${\rm CH_3\thinspace^{13}C}$ part of the molecule.  Initially, the spectrum appears similar to the 2-acetonitrile spectrum, with a doublet at $J$, and an additional doublet at $2J$ composed of several unresolved lines.  As the magnetic field is increased, additional lines in the multiplet at $2J$ become resolved.  At the highest magnetic fields, the spectrum displays the highest complexity, and is no longer recognizable from the perturbative treatment presented above.  The smooth trace at the top of the plot shows the log of the absorptive component of a high resolution numerical simulation, reproducing all features of the data, to the extent that lines are resolved.  Careful examination reveals 17 lines, 1 for the OH group and $(N+1)^2=16$ lines, as theoretically predicted in Ref. \cite{Appelt2010PRA}.


In conclusion, we have investigated near-zero-field nuclear magnetic resonance, where the effects of magnetic fields can be treated as a perturbation to the scalar $J$-couplings. This work represents a new form of NMR spectroscopy, complementary to high-field NMR, in which heteronuclear scalar couplings are almost always treated as a small perturbation to the dominant Zeeman interaction.  We find that the presence of small fields produces splitting of zero-field lines.  The splitting patterns have easy-to-understand rules and data are in excellent agreement with the predictions of first-order perturbation theory. It is interesting to note that the phenomenology observed here is similar to that of atomic spectroscopy of multi-electron atoms, and intuition developed in the latter field may be applied to interpretation of NZF NMR spectra. We have also investigated the case where Zeeman and $J$-couplings are comparable, resulting in signals with much higher complexity, potentially useful for NMR quantum computing \cite{Appelt2010PRA}.

This research was supported by the National Science Foundation under award \#CHE-0957655 (D. Budker and M. P. Ledbetter) and by the U.S. Department of Energy, Office of Basic Energy Sciences, Division of Materials Sciences and Engineering under Contract No. DE-AC02-05CH11231 (T. Theis, J.W. Blanchard, H. Ring, P. Ganssle and A. Pines).  We thank S. Knappe and J. Kitching for supplying the microfabricated alkali vapor cell.

\end{document}